\documentclass[11pt]{article}
\usepackage[dvips]{graphicx}

\begin{document}

\title{\Large \bf What Can Be Learned with a Lead-Based Supernova-Neutrino Detector?}
\author{ \\ J.\ Engel,$^{1}$  G.C.\ McLaughlin,$^{2}$ C.\ Volpe,$^{3,4}$ \\
{\small $^{1}$Department of Physics and Astronomy, University of North
Carolina,} \\ {\small Chapel Hill, NC 27599-3255 }\\
{\small $^{2}$Department of Physics, North Carolina State University,
Raleigh,
NC 27695-8202} \\
{\small $^{3}$ Institut de Physique Nucl\'eaire,
F-91406 Orsay cedex, France}\\
{\small $^{4}$ Institut f\"ur Theoretische Physik, Universit\"at
Heidelberg, Philosophenweg 19, D-69120 Heidelberg, Germany}}

\maketitle

\begin{abstract}
{\normalsize
We examine the prospects for using lead as a supernova-neutrino detector by
considering the spectrum of electrons produced, and the number of one- and
two-neutron events.  We show that the electron energy spectrum from
charged-current reactions can be used to extract information about the
high-temperature component of the neutrino spectrum. Some degree of
electron neutrino oscillation is expected in the supernova envelope. We
examine the prospects for untangling the signatures of various oscillation
scenarios, including, e.g.\ normal or inverted hierarchies, and different
values for the small mixing angle $\theta_{13}$.}

\end{abstract}

\section{Introduction}
\label{sec:intro}

The idea of detecting supernova neutrinos is exciting because although
galactic supernovae are rare, we could potentially learn a great deal from
the neutrinos.  Measuring their spectra would give us information about the
mass of the proto-neutron star core and its equation of state, and provide
input for supernova explosion calculations.  In addition, we can add to
what we've learned about neutrino oscillations from solar, atmospheric,
reactor, and accelerator neutrinos.  In this paper, we analyze the
capabilities of a detector based on lead \cite{omnis,land} that can see
both electrons from charged-current neutrino interactions and neutrons from
charged- or neutral-current interactions, e.g.\ a detector containing lead
perchlorate \cite{elliott}.

Supernova-neutrino detection differs from solar-neutrino detection.
Supernova neutrinos have a higher average energy, and an intense flux for a
very short period, $\sim 10 \, {\rm s}$.  Calculations of neutrino
diffusion in the proto-neutron star show that neutrinos emitted as $\mu$
and $\tau$ types have higher average energies, $\langle E_{\nu_{\mu,\tau}}
\rangle = 16 - 25 \, {\rm MeV}$, than those emitted as electron or
anti-electron types, which have $\langle E_{\nu_e}\rangle \sim 11 \, {\rm
MeV}$ and $\langle E_{\bar{\nu}_e} \rangle \sim 13 \, {\rm MeV}$
respectively.  The energy distributions can be fit to Fermi-Dirac spectra,
although some calculations show that they differ from Fermi-Dirac form on
their tails \cite{totani}.  This has been discussed extensively in
\cite{raffelt}.  
The interesting supernova physics conveyed by
neutrinos lies in the details of the energy distributions, acquired as the
neutrinos are emitted from the supernova core.  Neutrino oscillations will
mix the spectra in the outer envelope of the supernova.

Recent data from the Sudbury Neutrino Observatory (SNO) and Superkamiokande
(SuperK) have demonstrated that neutrinos oscillate, implying that they
have mass. SuperK indicates that atmospheric $\mu$ neutrinos oscillate into
objects that are not electron neutrinos, with $\delta m^2_{\rm atmos}
\approx 3 \times 10^{-3} {\rm eV}^2$ and a mixing angle of $\sin^2 2
\theta_{\rm atmos} \approx 1$ \cite{SuperK}.  Although it has long been
known that the flux of electron neutrinos from the sun is smaller than
expected, SNO has very recently used its sensitivity to neutral-current
scattering to measure the total flux from all species of neutrinos.  The
flux is approximately the same as predicted by the standard solar model
\cite{bahcall}, confirming that the electron neutrinos are oscillating into
some combination of $\mu$ and $\tau$ neutrinos \cite{SNO1}.  The signal
favors the large-mixing-angle (LMA) solution to the solar-neutrino problem,
corresponding to mixing parameters $\delta m^2_{\rm solar} \sim 10^{-5}
{\rm eV}^2$ and $\sin^2 2 \theta_{\rm solar} \approx 0.8$.  The LMA will be
tested by the reactor experiment KamLAND \cite{kamland}.

If all these results are cast in the form of 3-neutrino mixing, then there
is still an unknown mixing angle, which is currently limited by reactor
neutrino data to be \\
$\sin^2 2 \theta_{\rm reactor} 
\raisebox{-.25ex}{$\stackrel{<}{\scriptstyle \sim}$} 0.1$
\cite{chooz,paloverde}.  The LSND experiment \cite{lsnd} complicates the
picture, requiring a singlet neutrino or CPT violation in the neutrino
sector \cite{cptv}.
The oscillation parameters inferred from the solar and atmospheric results
imply that neutrinos will change flavor in the outer envelope of a
supernova.  The details of the transformation depend on {\it a)} the
unknown mixing angle $\theta_{\rm reactor}$ and {\it b)} whether there are
more than three species of neutrino, as implied by the combination of
atmospheric, solar and LSND results.

Several existing facilities can detect supernova neutrinos.  If a supernova
exploded 10 kpc from the earth with a luminosity of $3 \times 10^{53}$ ergs
and the energy partitioned equally among neutrino flavors, SuperK would see
 about 8300 events from $\bar{\nu}_e + p \rightarrow n + e^-$ in its water
detector \cite{beacom1}.  
KamLAND would see about 330 events from the same reaction in its
scintillator, and SNO would see about 360 such events in its light water
component \cite{beacom2}.  KamLAND may be able to measure the
spectrum of the high-temperature neutrinos through neutral-current
neutrino-proton elastic scattering \cite{beacom3}.
SNO would also be able to see 80 events from $\nu_e$
charged-current break-up of the deuteron. 
This number would increase in
the presence of oscillations.  An additional 500 events in SNO would come
from the neutral-current break-up of the deuteron \cite{vogel}, which has a
low threshold. An analysis of the supernova signal in a water-Cherenkov
detector and in a heavy-water detector has been conducted by in
ref. \cite{dutta}.
An analysis of neutrino mass limits from supernova time of flight 
is done in ref. \cite{beacom4}.

Here we consider a detector based on, e.g.\ lead perchlorate (${\rm
Pb(ClO_4)_2}$), first proposed by Elliott \cite{elliott}.  Lead had been
considered previously as a supernova detector by several groups, e.g.\
OMNIS \cite{omnis} and LAND \cite{land}.  Lead has an attractively large
neutrino-scattering cross section per nucleon compared with other elements,
and most of the scattering events produce neutrons \cite{fhm,kl,volpe}.
Lead perchlorate, which would be sensitive mainly to the higher-energy
neutrinos, has the appealing ability to measure energy deposited by the
electrons produced in charged-current reactions in coincidence with zero,
one, two, or more neutrons, as well as to measure the number of neutrons
emitted (in isolation) in neutral current reactions.  Per kt of lead
perchlorate, for the supernova described above, there would be about 378
electrons produced with one neutron and about 234 with two neutrons, if the
oscillations are induced only by the large solar mixing angle and the
high-energy neutrinos have a Fermi-Dirac distribution with temperature $T =
8.0~{\rm MeV},$ and effective chemical potential $\eta = 0$ (see eqn.\
(\ref{phi2})). In the neutral-current channel, there would be about 105
one-neutron events and about 72 two-neutron events. These numbers come from
the calculations described below.

In what follows, we investigate the supernova-neutrino signal in lead
perchlorate in more detail.  We discuss how the electrons produced by the
charged-current scattering on lead can be used to obtain spectral
information on the high-temperature neutrinos.  We then examine the
possibility that by using this information together with a comparison of
the numbers of charged- and neutral-current events, one could distinguish
among the oscillation scenarios discussed below.

In section \ref{sec:oscillations} we discuss the particulars of the
neutrino oscillations and what they would mean for the spectrum of electron
neutrinos coming from a supernova. In section \ref{sec:calc} we describe
our calculations of the cross sections for neutrino-induced spallation of
neutrons from lead via the charged and neutral currents.  We compare our
results to previous calculations and discuss uncertainties.  In section
\ref{sec:results} we present the results of our calculations and show how
they may be used to learn about the spectra and oscillation of supernova
neutrinos. Section \ref{sec:conclusions} is a conclusion.

\section{Oscillation Scenarios}
\label{sec:oscillations}

Although great strides have been made recently in understanding neutrino
oscillations, there are still unknown entries in the neutrino mixing
matrix. The mixing matrix can be written as
\begin{equation}
\left[\begin{array}{cc} \nu_e
\\ \\ \nu_\mu \\ \\ \nu_\tau \end{array}\right]
=
\left[\begin{array}{ccc}
c_{12} c_{13} & s_{12} c_{13} & s_{13} e^{-i \delta}
 \\ \\
-s_{12} c_{23} - c_{12} s_{23} s_{13} e^{i \delta} &
c_{12} c_{23} - s_{12} s_{23} s_{13} e^{ i \delta} &
s_{23} c_{13}
\\
s_{12} s_{23} - c_{12} c_{23} s_{13} e^{i \delta} &
-c_{12} s_{23} - s_{12} c_{23} s_{13} e^{i \delta} & c_{23} c_{13}
\end{array}\right]
\left[\begin{array}{cc} \nu_1
\\ \\ \nu_2 \\ \\ \nu_3 \end{array}\right]
\end{equation}
where $c_{ij}=\cos \theta_{ij}$ and $s_{ij}=\sin \theta_{ij}$,
$\theta_{ij}$ are the rotation angles, the subscripts $1,2,3$ denote the
mass eigenstates of the neutrinos, and $\delta$ is the CP-violating phase.
In combining the solar and atmospheric data into a single 3 by 3 mixing
matrix, one typically associates $\delta m^2_{21} = m_2^2 - m_1^2 \approx
\delta m^2_{\rm solar} \approx 10^{-5} {\rm eV}^2$. The mass-squared
difference associated with atmospheric neutrinos is $\delta m^2_{23}
\approx \delta m^2_{\rm atmos} \approx 3 \times 10^{-3} {\rm eV}^2$. The
relation $\sum \delta m^2 = 0$ makes it clear that $\delta m^2_{13}
\approx 3 \times 10^{-3} {\rm eV}^2$. What is not clear is the sign of
$\delta m^2_{23}$, since the vacuum oscillation of atmospheric neutrinos is
independent of the sign. As a consequence the sign of $\delta m^2_{31}$ is
unknown as well.  This has important implications for supernova neutrino
oscillations, since, depending on the last sign, either electron neutrinos
or electron antineutrinos will undergo a resonance associated with $\delta
m^2_{31}$. The MSW effect, unlike vacuum oscillations, depends on the sign
of the mass difference. The sign of $\delta m^2_{21}$ is determined by the
requirement that matter-enhanced oscillations in the sun cause a resonance
in the $\nu_e$ rather than the $\bar{\nu}_e$ channel.

By finding the vacuum survival probabilities of electron and muon
neutrinos, one can further associate $\sin^2 2 \theta_{\rm atmos} = \sin^2
2 \theta_{23} \cos^4 \theta_{13} \approx \sin^2 2 \theta_{23}$ and
translate the limit on the electron neutrino mixing angle from reactor
experiments to $\sin^2 2 \theta_{13} < 0.1$. Finally, one associates
$\sin^2 2 \theta_{\rm solar}$ with $\sin^2 2 \theta_{12}$.

What are the consequences of the above analysis for supernova neutrinos? In
the 3 by 3 mixing scheme there are two densities at which resonances can
occur \cite{dighe}.  In some cases, shock propagation \cite{shirato} and
nonaxisymmetric explosions \cite{blondin} will cause the neutrinos to
encounter their resonance density more than once and thus cause
time-dependent features in the profile.  Here we consider the problem in
the first couple of seconds after the explosion, before the shock has had a
chance to propagate to the point where the neutrinos may undergo a
particular resonance twice.

At these relatively early times, there are two main resonances through
which the electron neutrinos will propagate.  The densities at which the
resonances occur are set by the $\delta m^2$'s and the energies of the
neutrinos.  For a given density, the energy that a neutrino must have to
undergo a resonant transformation is given by
\begin{equation}
 \label{eres}
 E_{\rm res}(r) \equiv \pm\frac{\delta m^2 \cos 2\theta_v}{V(r)},
\end{equation}
where the potential is
$ V(r) \equiv 2 \sqrt{2} G_F
             \left[N_{e^-}(r) - N_{e^+}(r)\right]$.

\begin{figure}[ht]
\centerline{\includegraphics[angle=0,width=10cm]{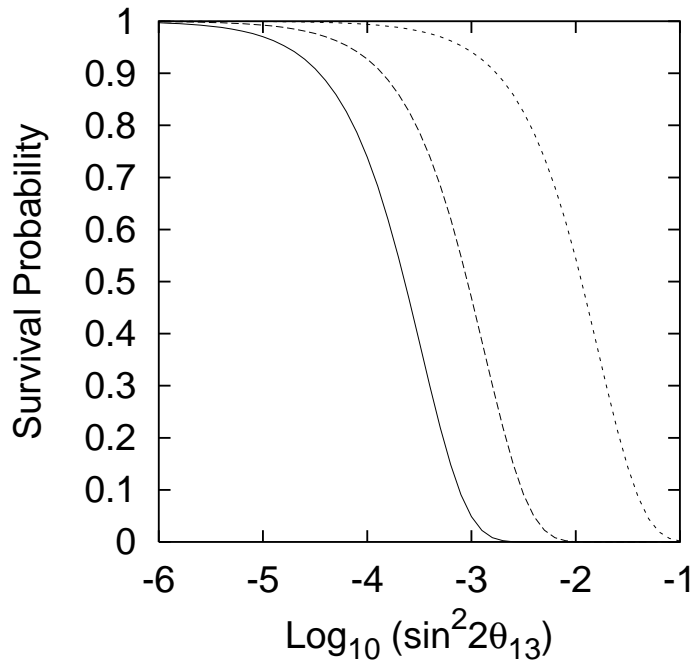}} \caption{\small
Survival probability as a function of $\sin^2 2 \theta_{13}$ for an
electron neutrino with energy 10 MeV, 40 MeV and 100 MeV (left to right),
as it goes through the resonance associated with $\delta m^2_{13} \approx 3
\times 10^{-3} {\rm eV}^2 $.  For values below $\sin^2 \theta_{13} \sim
10^{-3}$ the neutrinos above 40 MeV pass by unaffected, while for values
much larger than this, the neutrinos convert. The neutrinos that do not
convert in this resonance will partially convert at the later resonance
associated with $\delta m^2_{21}$.} \label{fig:trans}
\end{figure}

\underline{The $\delta m^2_{13}$ resonance:} The sign of $\delta m^2_{\rm
solar}$ is fixed by the solution to the solar neutrino problem.  However,
as noted above, the sign of $\delta m^2_{13}$ is still unknown.  In a
``normal hierarchy" ($\delta m^2_{13}>0$), electron neutrinos, as opposed
to antineutrinos, would undergo a resonance at a density of about $10^3 \,
{\rm g} \, {\rm cm^{-3}}$. (This is for a 40 MeV neutrino and a mass scale
$\delta m^2_{13} \approx 3 \times \, 10^{-3} \, {\rm eV}^2$.) If the
resonance were adiabatic, which would be the case if $\sin^2 2 \theta_{13}
\gg 10^{-3}$ (see fig.\ \ref{fig:trans}), then the original electron
neutrinos would become mostly $\nu_3$, which would be measured in the
detector as part $\nu_\mu$ and part $\nu_\tau$. The original $\nu_{e}$
would therefore be seen primarily in the neutral current events. The
electron neutrinos would almost all acquire the high-temperature spectrum
of the original $\nu_\mu$ or $\nu_\tau$. The $\nu_3$ would not mix with the
other species again in the supernova, so the electron neutrinos would
retain the ``hot'' spectrum as they arrive at the earth.

\underline{The $\delta m^2_{12}$ resonance :} If the electron neutrinos
didn't transform at the density associated with $\delta m^2_{13}$, which
would imply $\sin^2 2 \theta_{13} \ll 10^{-3}$, they would have a second
opportunity at a lower density, $\sim 1 \, {\rm g} \, {\rm cm}^{-3}$, which
is associated with $\delta m^2_{\rm solar} \approx 10^{-5} {\rm eV}^2$.  We
assume that this resonance is adiabatic, as implied by the presupernova
model profiles of \cite{woosley}.  Then the electron neutrinos as they
would be measured as they arrive at the earth would have $\sin^2
\theta_{\rm solar}$ of the original $\nu_e$ spectrum together with $\cos^2
\theta_{\rm solar} $ of the original $\nu_{\mu,\tau}$ spectrum. The number
of charged-current events in the detector would therefore be $\sin^2
\theta_{\rm solar}$ times the number of events expected with only the
lower-temperature spectrum and $\cos^2 \theta_{\rm solar}$ times the number
of events expected with the higher-temperature spectrum.  The spectrum
could be further modified by the earth-matter effect if the neutrinos
travelled through the earth before reaching the detector.

We do not consider scenarios intermediate between those just described,
i.e. a partially adiabatic first resonance, with partial transformation. In
other words, as fig.\ \ref{fig:trans} shows, we are considering situations
where $\sin^2 2 \theta_{13} \gg 10^{-3}$ or $\sin^2 2 \theta_{13} \ll 10^{-3}$.

\underline{Inverted hierarchy:}   So far we have assumed a normal hierarchy
where the third mass eigenstate is the heaviest.  An ``inverted'' hierarchy
occurs when this state is the lightest.  This would cause a resonance to
occur in the electron {\it anti}neutrino channel at the higher density
resonance associated with $\delta m_{13}^2$.  Since a lead-based detector
sees very few $\bar{\nu}_e$'s, the inverted hierarchy would produce nearly
the same signal from lead in OMNIS or LAND as the normal hierarchy with
very small $\theta_{13}$.

We limit our discussion to the situation where there is relatively little
earth matter effect and are no sterile neutrinos.

\section{Calculations}
\label{sec:calc}

In this work we assume a Fermi-Dirac distribution for the high-energy
neutrino component.  The number flux of neutrinos is given by
\begin{equation}\label{phi1}
d\Phi_\nu(E_\nu) =
{L_\nu \over 4 \pi r^2} {1 \over \langle E_\nu \rangle} f(E_\nu) d E_\nu
\end{equation}
where
\begin{equation}\label{phi2}
f(E_\nu) = {1 \over T_\nu^3 F_2(\eta)}
{E_\nu^2 \over \exp(E_\nu/T_\nu  - \eta) + 1}
\end{equation}
 and
$\langle E_\nu \rangle = T_\nu F_3(\eta)/F_2(\eta)$.
The functions $F_2$ and $F_3$ are the Fermi integrals
\begin{equation}
\label{phi3} F_n = \int_0^\infty {x^n dx \over \exp(x  - \eta) +
1}
\end{equation}
We will assume equipartition of energy between the six neutrinos and
antineutrinos, and vary the values of $T$ and $\eta$.  Because
the total energy in each species is fixed, if the average neutrino energy
rises, the number flux of neutrinos decreases.

To calculate the event rate from supernova neutrinos we also need to know the
cross sections of the charged- and neutral-current reactions of
interest :
\begin{eqnarray}
\nu_e + ^{208}{\rm Pb} &\rightarrow& ^{207}{\rm Bi} + n + e^-\\
\nu_e + ^{208}{\rm Pb} &\rightarrow& ^{206}{\rm Bi} + 2 n + e^-\\
\nu_x + ^{208}{\rm Pb} &\rightarrow& ^{207}{\rm Pb} + n \\
\nu_x + ^{208}{\rm Pb} &\rightarrow& ^{206}{\rm Pb} + 2 n
\end{eqnarray}
 Of course, only half of natural lead
is $^{208}$Pb, but the cross sections from the other isotopes
should not be very different 
because the thresholds are the same to within a couple of MeV and the
strength distributions depend largely on collective nuclear motion that is not much
affected by the addition or removal of a few nucleons.  The reaction rates in the
other isotopes 
can be calculated through slight modifications to the methods described
below.

  Even in $^{208}$Pb, these cross sections have not yet been
  measured. 
We
therefore must rely on theoretical predictions.  The cross section for a neutrino
with energy $E_{\nu}$ to create an electron with energy $E_e$ is
\begin{equation}
\sigma(E_{\nu},E_e)={G^{2} \over {2 \pi}} \cos^{2}\theta_C p_eE_e
\delta(E_{\nu}-E_e-E_{fi}) \int_{-1}^{1}d(\cos \, \theta)M_{\beta} ~,
\label{e:1}
\end{equation}
where $G \, \cos \, \theta_C$ is the effective weak coupling constant,
$\theta$ is the angle between the directions of the incident neutrino and
the outgoing electron, $p_{e}$ is the outgoing electron momentum, and
$E_{fi}$ is the excitation energy of the $^{208}$Bi with respect to the
ground state of $^{208}$Pb. The quantity $M_{\beta}$ is a combination of
matrix elements operators of the form $j_l(qr) Y_l(\Omega) \tau_+\,$,
$j_l(qr) Y_l(\Omega) \vec{\sigma} \tau_+\,$, $j_l (qr) Y_l(\Omega )
\vec{\nabla} \tau_+\,$, and $j_l (qr) Y_l(\Omega) \vec{\sigma} \vec{\nabla}
\tau_+$ (these include the usual Fermi and Gamow-Teller operators in the
long-wavelength approximation) where $\tau_+$ changes a neutron into a
proton and $q\equiv|\vec{q}|$ is the momentum transferred to the nucleus,
which ends up as $^{208}$Bi.

Our calculations are based on the coordinate-space Skyrme-Hartree-Fock
method in a 20 fm box, followed by the $A$- and $B$-matrix version of RPA
with the same Skyrme interaction in the basis of Hartree-Fock states. We
actually do the calculations twice, using two different sets of expressions
(\cite{Walecka,Kubo}) for $M_{\beta}$ as well as two different Hartree-Fock
and RPA codes. 
For $T=8$, $\eta=0$ neutrinos, the cross sections from the two versions
differ only by 8\% 
when one neutron is emitted and 4\% when two are emitted. We
include
all multipoles up to $J=4$.  (The contributions of higher multipoles are
smaller at these neutrino energies.) We truncate the Hartree-Fock spectrum
at a value high enough to contain most of the strength; increasing it by a
few MeV produces almost no change. In particular, the configuration space
used is large enough to satisfy both the non-energy-weighted and the
energy-weighted sum rules. We reduce the value of  the axial-vector
coupling constant $g_A$ from 1.26 to 1 in the $1^+$ channel to account for
spin-quenching (this may still be a slight overestimate for lead; see ref.\
\cite{fhm}). Finally, we treat the Coulomb interaction of the outgoing
electron with the final nucleus as in ref.\ \cite{volpe}. (We comment on
the procedure for the Coulomb corrections below.)

\begin{figure}[hbt]
\centerline{\includegraphics[angle=0,width=10cm]{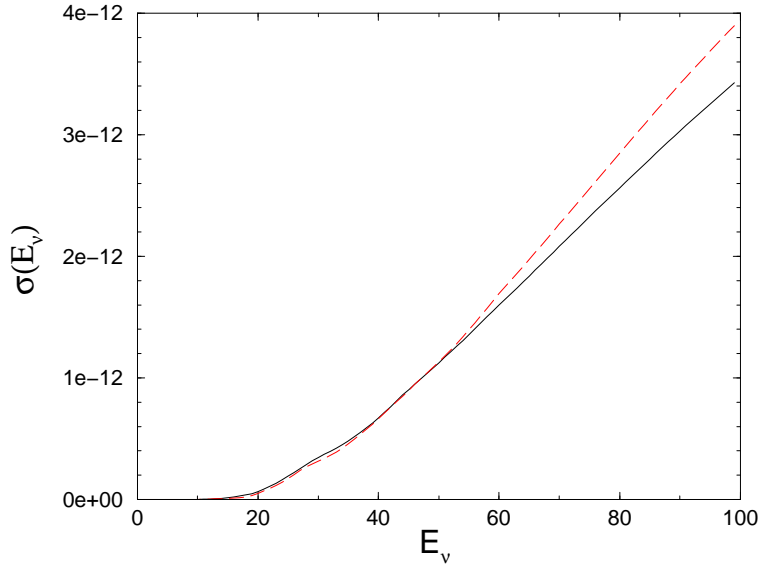}}
\caption{\small Total charged-current cross sections, in fm$^2$ as a
function of neutrino energy, for the Skyrme interactions SIII (solid line)
and SkO+ (dashed line). \label{fig:enu}}
\end{figure}

An amount of energy $E_{fi}=E_{\nu}-E_e$ is transferred to the nucleus, so
that a number of nuclear states can be excited. Much of the cross section
is in giant resonances, e.g.\ the Gamow-Teller (GT) resonance, 15.6 MeV
above the ground state of $^{208}$Bi, and the dipole and spin-dipole
resonances, which make up a broad peak centered about 25 MeV above the
ground state. It's important for calculations to reproduce these
resonances; an error of a few MeV can make about 20$\%$ difference in
 the flux-averaged
cross sections \cite{volpe}.  In our RPA calculations we used two 
Skyrme forces, the
venerable SIII \cite{s3} and the force SkO' \cite{sko'} with the
``time-odd" terms in the energy functional explicitly adjusted to reproduce
GT strength distributions as described in ref.\ \cite{Bender}; we refer to
the latter as SkO+.  The interactions both correctly locate the isobaric
analog state and GT centroid to within an MeV, put the dipole and
spin-dipole strength in the right region, and give similar results more
generally.  Fig.\ \ref{fig:enu} shows the total cross sections from the two
as a function of energy.  They are nearly identical except at high energies
that contribute little to the supernova-neutrino flux.

For neutral current cross sections, eq.\ (\ref{e:1}) still holds provided
the outgoing electron is replaced by a neutrino or antineutrino (the
quantity $M_{\beta}$ contains different coefficients, and isn't corrected
for Coulomb effects).  We again quench all matrix elements in the $1^+$
channel by $(1/1.26)^2$, and assume that the nucleon  carries no isoscalar
(i.e.\ strange) axial current.  Our calculation here follows the same steps
as in the charged-current channel except that we use the like-particle RPA
instead of the charge-changing RPA.

We want to calculate the individual cross sections for the emission of 0,
1, or 2 neutrons.  Following ref.\ \cite{fhm}, we do so by assuming that
two or more neutrons are emitted if the excitation energy of the final
nucleus is more than 2.2 MeV above two-neutron-emission threshold, and one
neutron if it's less than that but greater than the 100 keV above
one-neutron-emission threshold.  For high-energy supernova neutrinos, the
$1^+$ excitations in charged-current reactions are usually associated with
the emission of one neutron and the dipole and spin-dipole states with
the emission of two. For example, with a neutrino flux characterized by
$T=8$, $\eta =0$ --- typical predicted values for the hot
spectrum --- the $1^+$ states contribute 53\% of the one-neutron cross
section and only 5\% of the two neutron cross section.  The dipole and
spin-dipole states ($0^-$, $1^-$, and $2^-$) contribute 62\% of the
two-neutron cross section but only 15\% for one-neutron emission.

Our results are nearly identical with those in ref.\ \cite{volpe} and are
close to those of ref.\ \cite{kl}; the total cross sections as a function
of energy agree to within $20-30\%$ in the charged-current channel and 15\%
in the neutral-current channel.  They are considerably smaller, however
than those of ref.\ \cite{fhm}, as noted in refs.\ \cite{kl,volpe}.
Although ref.\ \cite{fhm} uses a different nuclear model, the reasons for
the difference in cross sections are more basic.  Ref.\ \cite{volpe}
pointed to the neglect of terms beyond lowest order in the long-wavelength
expansions of the multipole operators in $M_{\beta}$, and to a different
treatment of Coulomb corrections. The former increases the cross sections
(compared to ours) dramatically, while the latter reduces them somewhat,
but still leaves them too large.

We have treated the Coulomb corrections the same way as in ref.\
\cite{volpe}:  at each electron energy we take the smaller of the results
with a simple Fermi function and the effective momentum approximation
\cite{ema}. This procedure almost certainly overestimates the cross section
a little as well, but not badly \cite{Engel,volpe}.  We also tried the
procedure of ref.\ \cite{fhm}: assuming that the outgoing electron was
always in an s-wave and correcting its wave function according to the
tables of ref.\ \cite{Janecke}, which contain the effects of a Fermi
function for a spherical charge distribution with nonzero radius.  This
prescription, which typically underestimates Coulomb effects at high
electron energies, reduces the total cross section at $T=8$, $\eta=0$ by
about 20\%.   We give our results in tabular form (with the first Coulomb 
prescription and the force SIII) in
Table \ref{table:xsec}.

\begin{table}
\caption{Neutrino cross sections in units of $10^{-40}$ cm$^2$ as a
  function of energy (MeV)
for emission of one and two neutrons, and summed over all decay channels, obtained
with 
the Skyrme force SIII.  We include the charged-current channel for neutrinos, and the
neutral-current channel for both neutrinos and
antineutrinos. \label{table:xsec} }
\begin{center}
\begin{tabular} {|c||c|c|c||c|c|c||c|c|c|}
\hline
$E_{\nu}$ & \multicolumn{3}{c|}{$\nu_e \rightarrow e$} & 
\multicolumn{3}{c|}{$\nu \rightarrow \nu$} &
\multicolumn{3}{c|}{$\bar{\nu} \rightarrow \bar{\nu}$}  \\  \cline{2-10}
 & 1n & 2n & total & 1n & 2n & total & 1n & 2n & total \\ \hline
 5   & --- & --- & 0.39E-07 & --- & --- & 0.67E-11 & --- & --- & 0.66E-11  \\
10   & 0.29E-11 & --- & 0.09 & 0.002 & --- & 0.007 & 0.002 & --- & 0.007  \\
15   &    0.91 &    --- &    1.54 &    0.06 &    --- &    0.08 &    0.05 &    --- &   
0.08  \\
20   &    4.96 &    --- &    6.51 &    0.20 &    --- &    0.27 &    0.18 &    --- &   
0.24  \\
25   &   14.66 &    0.45 &   17.63 &    0.46 &    0.03 &    0.62 &    0.40 &    0.03 &   
0.54  \\
30   &   25.05 &    3.15 &   32.22 &    0.87 &    0.15 &    1.22 &    0.73 &    0.13 &   
1.04  \\
35   &   29.27 &   10.85 &   45.37 &    1.44 &    0.42 &    2.15 &    1.18 &    0.36 &   
1.79  \\
40   &   33.56 &   23.68 &   64.10 &    2.15 &    0.93 &    3.48 &    1.73 &    0.76 &   
2.82  \\
45   &   37.91 &   38.97 &   85.33 &    2.97 &    1.74 &    5.25 &    2.34 &    1.39 &   
4.17  \\
50   &   42.54 &   53.79 &  106.16 &    3.86 &    2.93 &    7.50 &    2.99 &    2.26 &   
5.82  \\
55   &   47.17 &   71.63 &  130.09 &    4.79 &    4.56 &   10.24 &    3.65 &    3.42 &   
7.78  \\
60   &   52.02 &   90.05 &  154.64 &    5.74 &    6.63 &   13.50 &    4.31 &    4.85 &  
10.04  \\
65   &   56.31 &  108.73 &  178.75 &    6.71 &    9.17 &   17.25 &    4.97 &    6.54 &  
12.57  \\
70   &   60.39 &  129.14 &  204.17 &    7.69 &   12.17 &   21.49 &    5.62 &    8.47 &  
15.34  \\
75   &   64.03 &  150.40 &  229.88 &    8.67 &   15.59 &   26.14 &    6.25 &   10.62 &  
18.31  \\
80   &   67.04 &  170.75 &  253.92 &    9.65 &   19.39 &   31.16 &    6.86 &   12.94 &  
21.42  \\
85   &   69.69 &  191.16 &  277.58 &   10.58 &   23.51 &   36.43 &    7.44 &   15.39 &  
24.61  \\
90   &   71.95 &  211.73 &  300.95 &   11.45 &   27.90 &   41.88 &    7.97 &   17.93 &  
27.82  \\
95   &   73.91 &  231.25 &  323.03 &   12.23 &   32.47 &   47.39 &    8.45 &   20.51 &  
31.00  \\
\hline
\end{tabular}
\end{center}
\end{table}

How accurate are our calculations?  The good agreement between the two
quite different Skyrme forces and between our calculations and those of
refs.\ \cite{kl,volpe} is encouraging.  It is possible, however, that all
such calculations contain similar mistakes. They are all done in the RPA,
and they all treat Coulomb effects in a similar way.  There may be
systematic nuclear structure uncertainties, related, e.g., to the issue of
quenching of forbidden multipoles in charge-exchange reactions, that no
data accurately address. It's hard to assert, especially without a more
detailed comparison to available data from hadronic charge-exchange and
electron-scattering experiments, that the cross sections are accurate to
better than 25\% for high-temperature supernova-neutrinos.  The average
recoil energy is a much more robust quantity, however; there our two
calculations typically differ by less than a few percent, and even when we
change the treatment of Coulomb effects, the difference is typically less
than 5 or 10\%.  We can probably take that range as a reasonable estimate
of the uncertainty in the average energy. We will make use of this
observable in the next section to show how the temperature of the supernova
neutrinos can be significantly constrained.

\section{Results}
\label{sec:results}

Here we use the calculations described in the previous sections to show: 1)
how to use charged-current events in which an electron is emitted in
coincidence with one neutron or with two neutrons to characterize the
neutrino spectrum, and 2) how to use the charged- and neutral-current
signal to determine what sort of oscillation has taken place.

\begin{figure}[htb]
\begin{center}
\begin{minipage}{7.5cm}
\centerline{\includegraphics[angle=0,width=6cm]{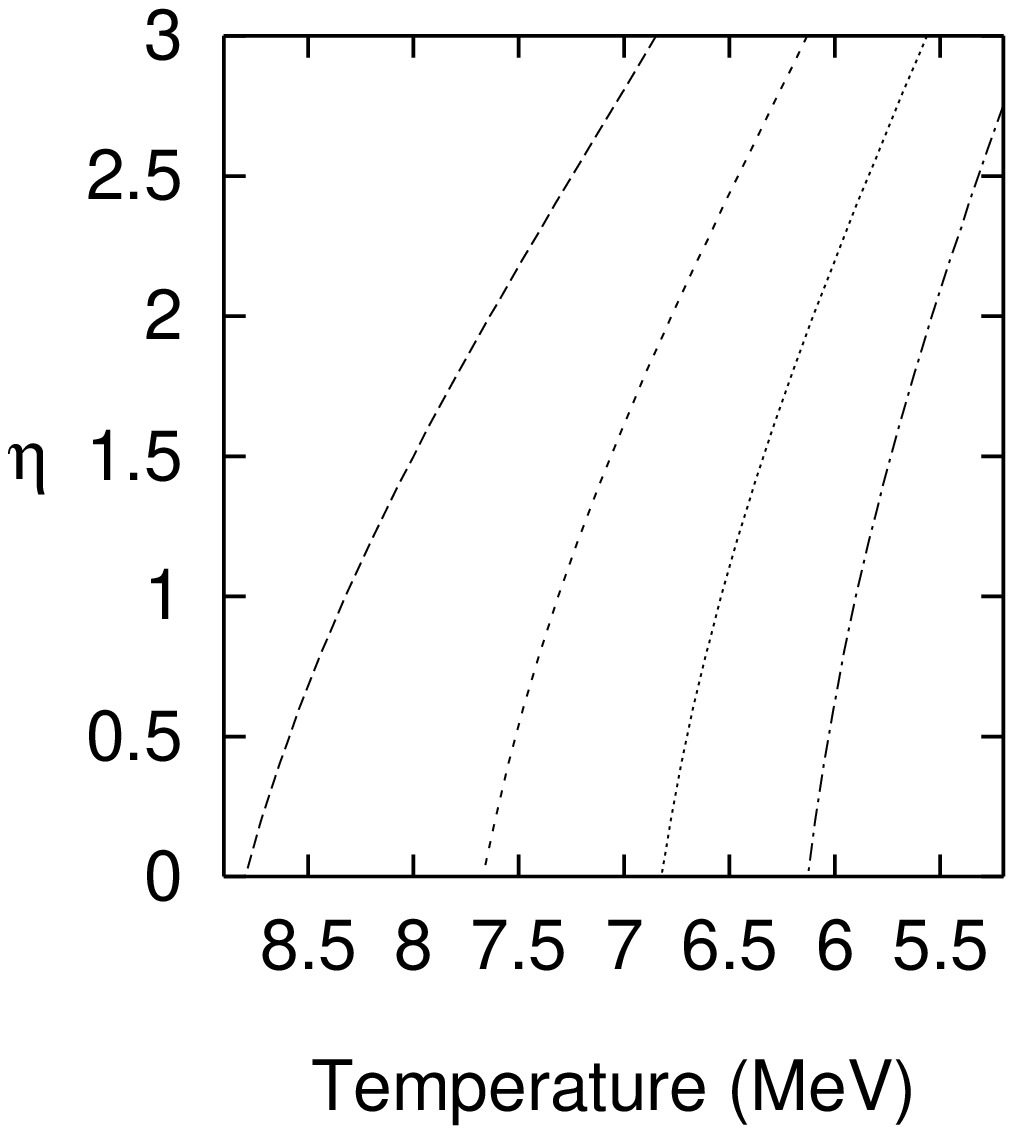}}
\end{minipage}\begin{minipage}{7.5cm}
\centerline{\includegraphics[angle=0,width=6cm]{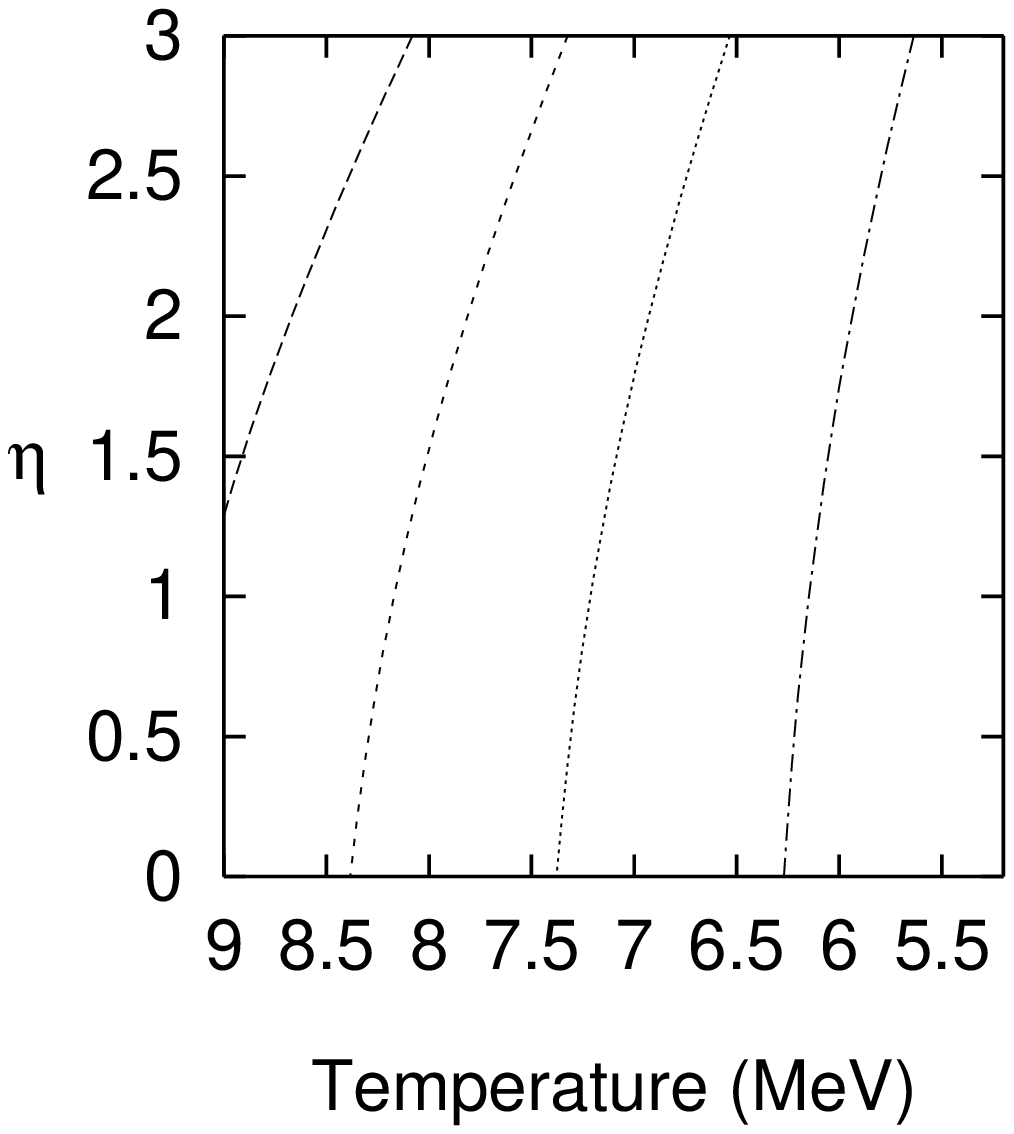}}
\end{minipage}
\end{center}
\caption{\small These contour plots show the number of electrons emitted
from one kilotonne of lead with one neutron (left panel) and two neutrons
(right panel).  The contours on the left panel are from left to right 500,
450, 400, and 350.  The contours on the right panel are from left to right
450, 350, 250, and 150. The horizontal axis is the temperature of a
Fermi-Dirac distribution of neutrinos, while the vertical axis shows the
effective chemical potential. \label{fig:numbers}}
\end{figure}

\begin{figure}[t!]
\centerline{\includegraphics[angle=0,width=10cm]{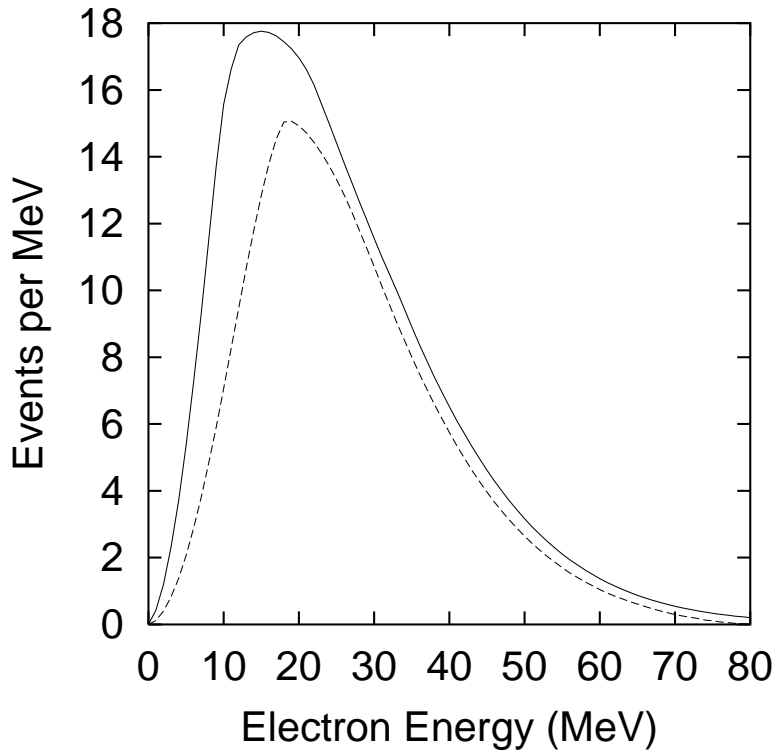}} \caption{\small
Energy spectra of electrons associated with one-neutron (solid line), and
two-neutron (long dashed line) events obtained by folding the cross
sections (Eq. \ref{e:1}) with the neutrino fluxes (Eq. \ref{phi1}) for a
neutrino spectrum parameterized by $T = 8$ MeV and  $\eta = 0$ in one
kilotonne ($10^6 {\rm kg}$) of Pb. \label{fig:espect}}
\end{figure}

First, we display in fig.\ \ref{fig:numbers} the numbers of events
associated with various parameterizations of Fermi-Dirac spectra.  Next we
consider the electron energies.  It takes a neutrino of about 18 MeV to
excite lead with much probability. The observed electron energies are
therefore chiefly sensitive to the high-energy tail of the neutrino
spectrum.  In fig.\ \ref{fig:espect} we show the electron-energy spectra
associated with one and two neutrons, obtained by folding the cross
sections calculated with the Skyrme force SIII (which we use below unless
we explicitly state otherwise) with the neutrino flux in eqs.\
(\ref{phi1}-\ref{phi3}) for $T=8$ MeV, $\eta = 0$. The spectra observed in
the detector will have very nearly these shapes with either partial or
complete oscillations. The high thresholds associated with one- and
(especially) two-neutron emission make the signal even less sensitive to
the original low-temperature component of the spectrum.

If there were no oscillations at all, and all the electron neutrinos had a
temperature $T=3\,{\rm MeV}$ and an effective chemical potential $\eta_{\rm
eff}=3$, there would be only $\sim$ 92 one-neutron events and $\sim 5$
two-neutron events. If oscillations were to occur, these numbers would be
much larger. If the transformation associated with the $\delta m^2_{13}$
were complete, there would be $\sim 460$ one-neutron events, with an
average electron energy of 22.9 MeV, and $\sim 310$ two-neutron events,
with an average energy of 26.2 MeV.  If instead the oscillations were due
to the resonance associated with $\delta m^2_{12}$ ($\sin^2 2 \theta_{\rm
solar} \approx 0.75$), the $\sim 370$ one-neutron events would have an
energy of 22.1 MeV, hardly any different than before, and the two-neutron
events would simply be reduced in number by almost exactly 25\%, with
little change in the spectral shape.

\begin{figure}[t]
\begin{center}
\begin{minipage}{7.5cm}
\centerline{\includegraphics[angle=0,width=6cm]{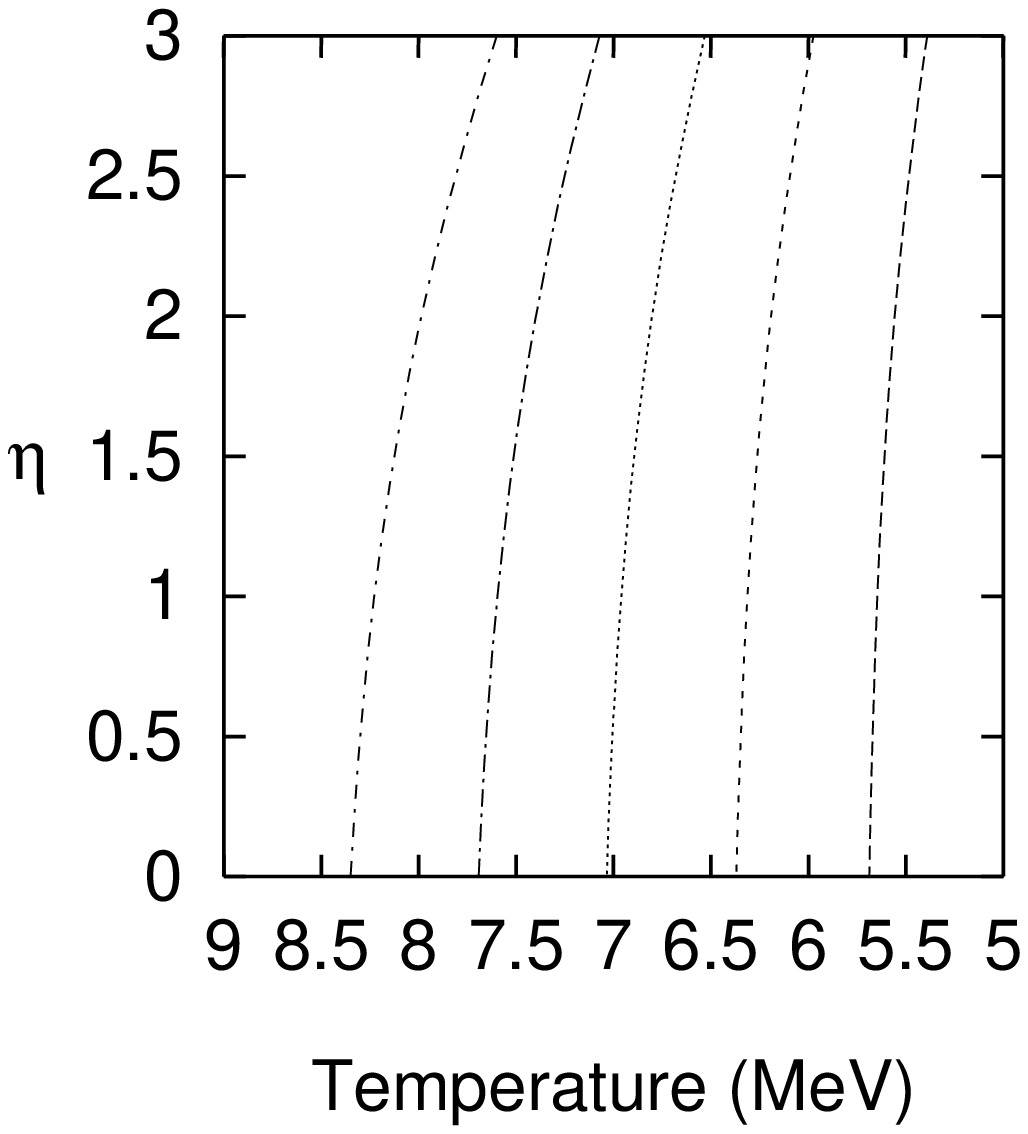}}
\end{minipage}\begin{minipage}{7.5cm}
\centerline{\includegraphics[angle=0,width=6cm]{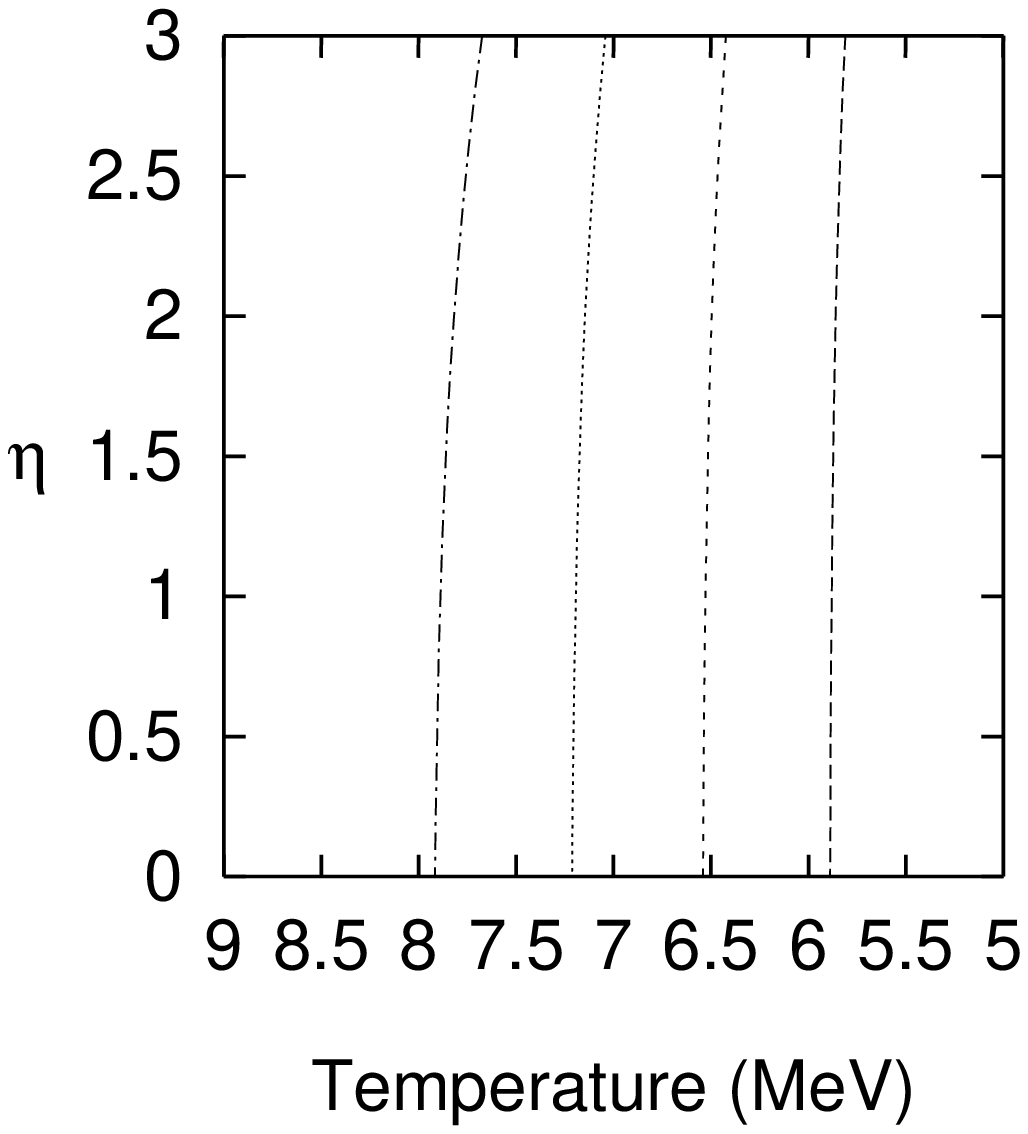}}
\end{minipage}
\end{center}
\caption{\small These contour plots show the average energies of electrons
emitted from lead with one neutron (left panel) and two neutrons (right
panel). The contours on the left panel are from left to right 24 MeV, 22
MeV, 20 MeV, 18 MeV, and 16 MeV while those on the right panel are from
left to right 26 MeV, 24 MeV, 22 MeV, and 20 MeV. \label{fig:e2}}
\end{figure}

Because the shapes of the electron-energy spectra do not depend
significantly on whether the oscillations are complete or partial, they can
provide a lot of information on the original $\mu$/$\tau$-neutrino
spectrum.  Of course, we do not know the cross sections perfectly, so
extracting the neutrino spectrum from the electron spectra is not trivial.
We therefore focus on an observable that proved fairly robust against
changes in the calculation: the average electron energy for one- and
two-neutron events. In fig.\ \ref{fig:e2} we show a few contours in a plot
of the average energy as a function of the temperature and effective
chemical potential, assuming a Fermi-Dirac distribution for the
higher-energy component of the neutrinos.  The figure shows that the higher
the temperature of the neutrinos, the higher the average electron energy.
The same is true of the chemical potential, particularly when it reaches 2
or 3.

\begin{figure}[b!]
\begin{center}
\begin{minipage}{7.5cm}
\centerline{\includegraphics[angle=0,width=6cm]{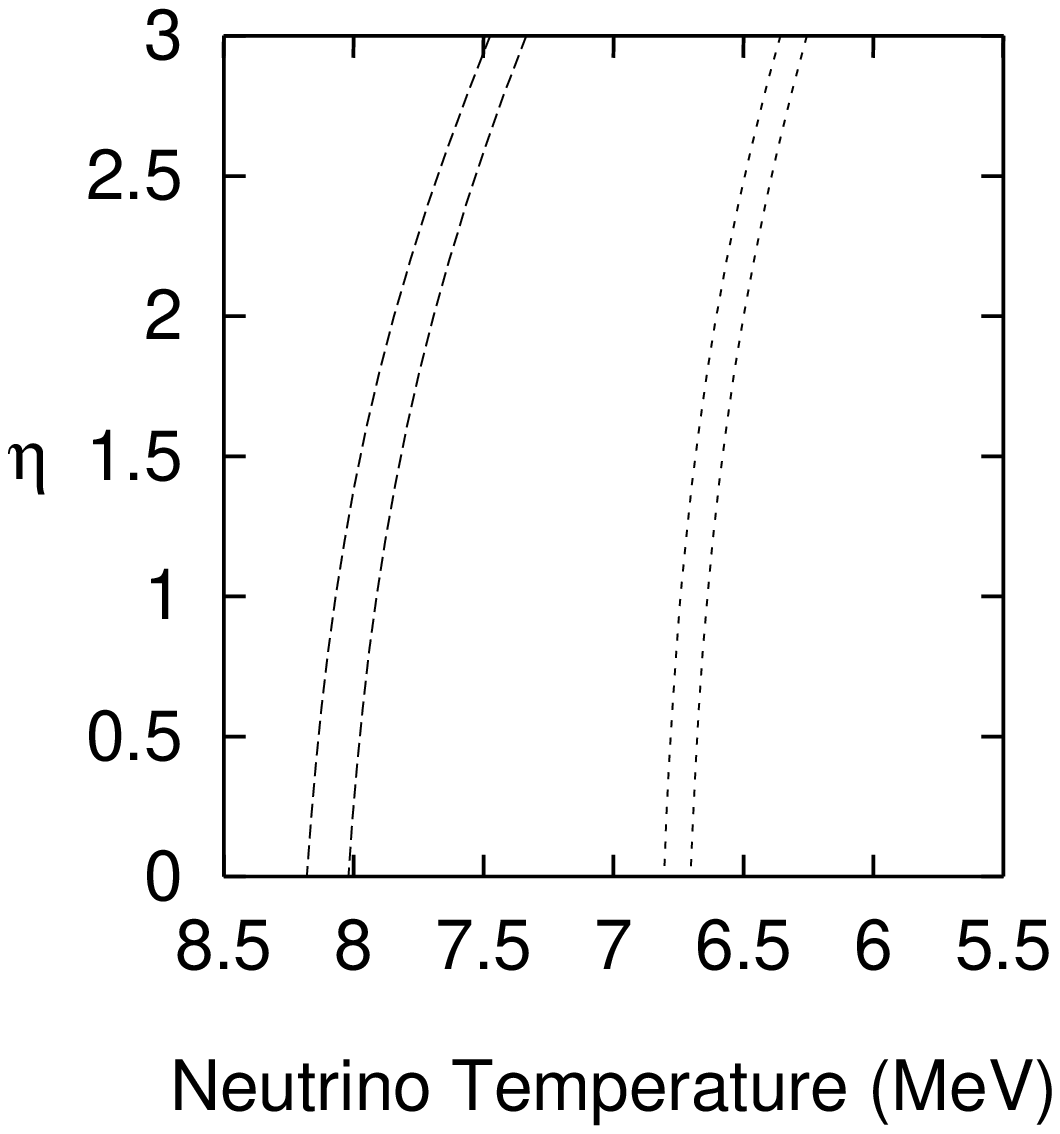}}
\end{minipage}\begin{minipage}{7.5cm}
\centerline{\includegraphics[angle=0,width=6cm]{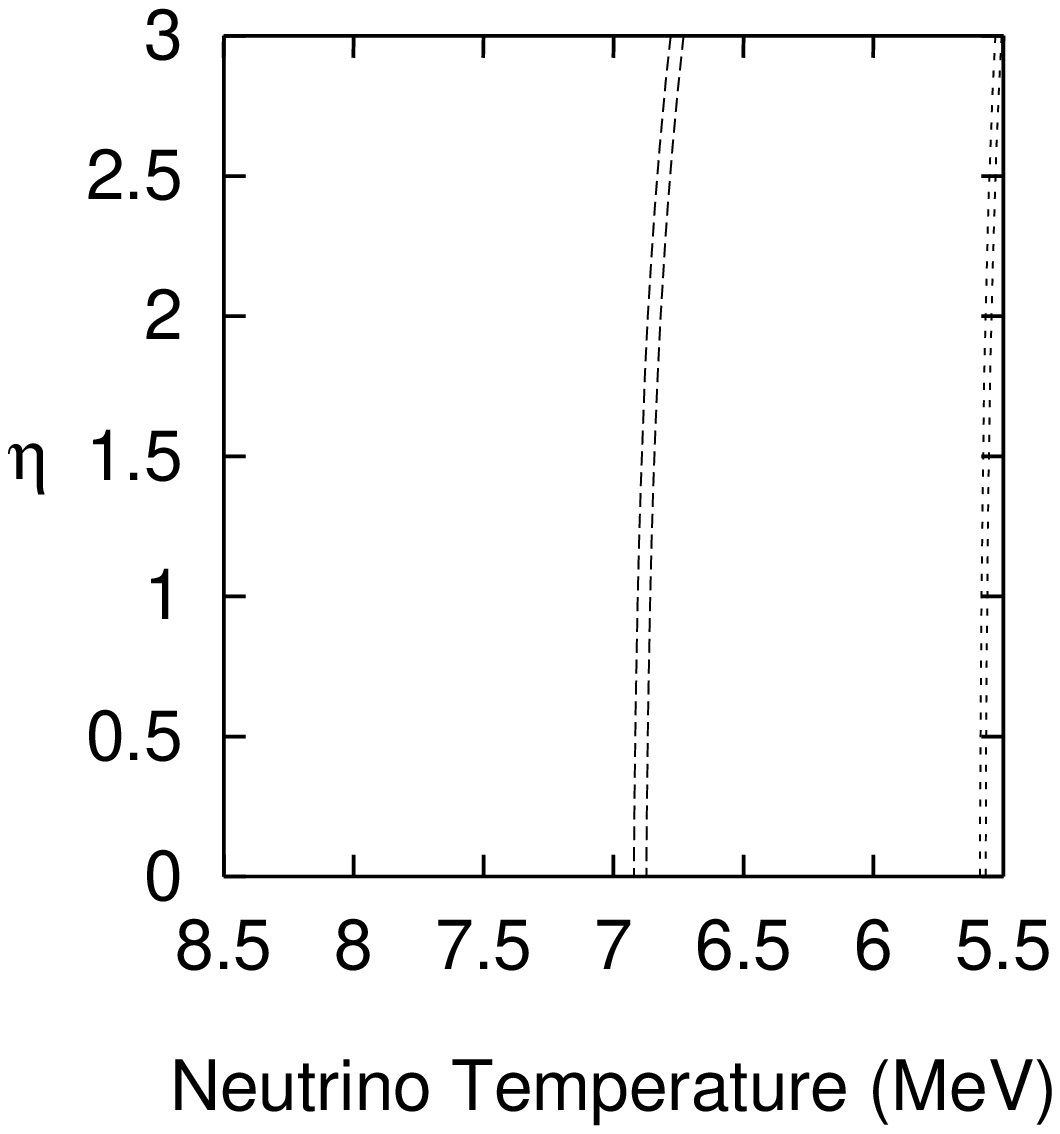}}
\end{minipage}
\end{center}
\caption{\small These contour plots show the average energies of electrons
emitted from lead with one neutron (left panel) and two neutrons (right
panel). The two contours on the left (right) side of each panel are for 23
Mev (19 MeV) average energy for two different forces. \label{fig:e1}}
\end{figure}

\begin{figure}[t!]
\begin{center}
\begin{minipage}{7.5cm}
\centerline{\includegraphics[angle=0,width=6cm]{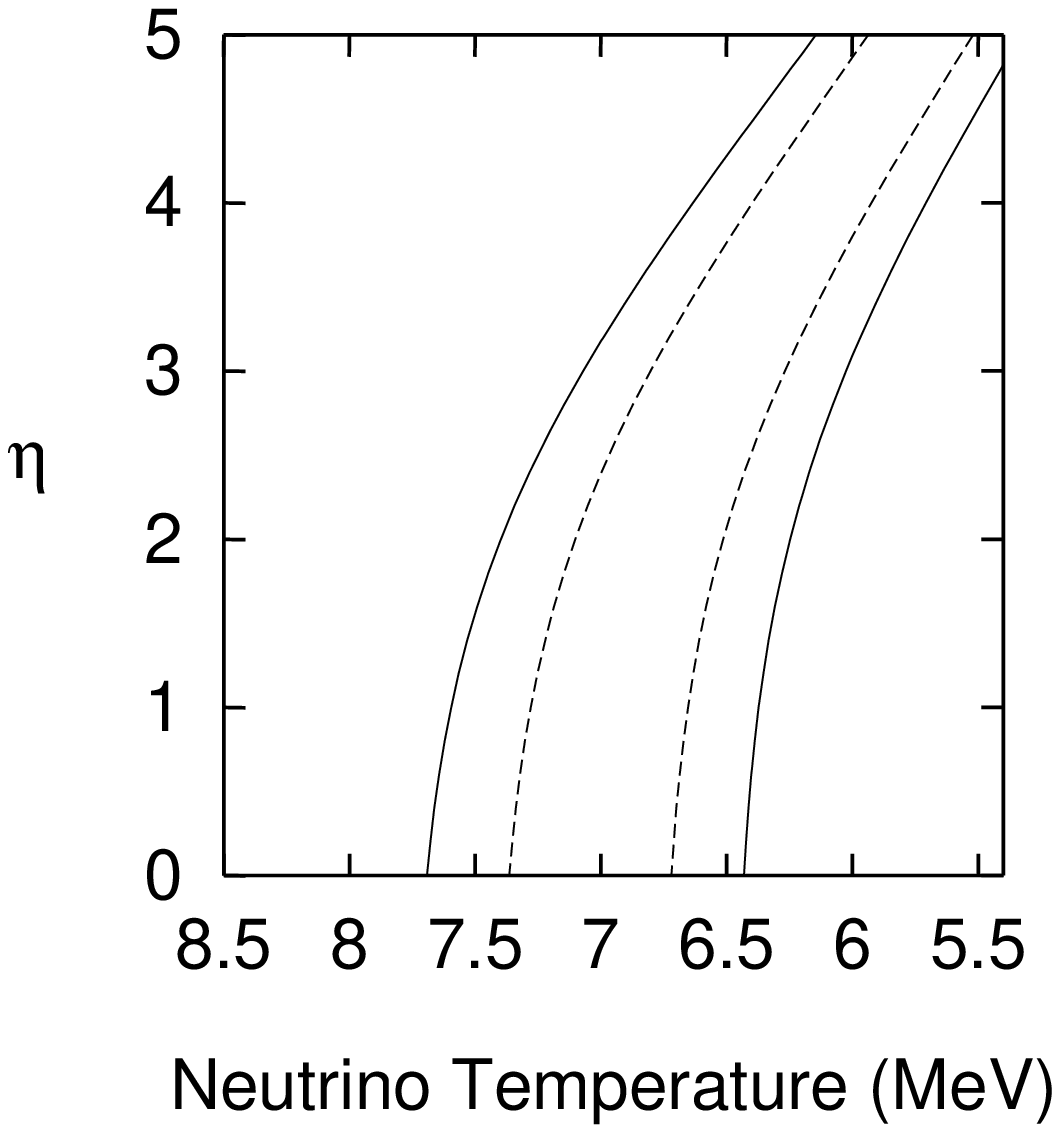}}
\end{minipage}\begin{minipage}{7.5cm}
\centerline{\includegraphics[angle=0,width=6cm]{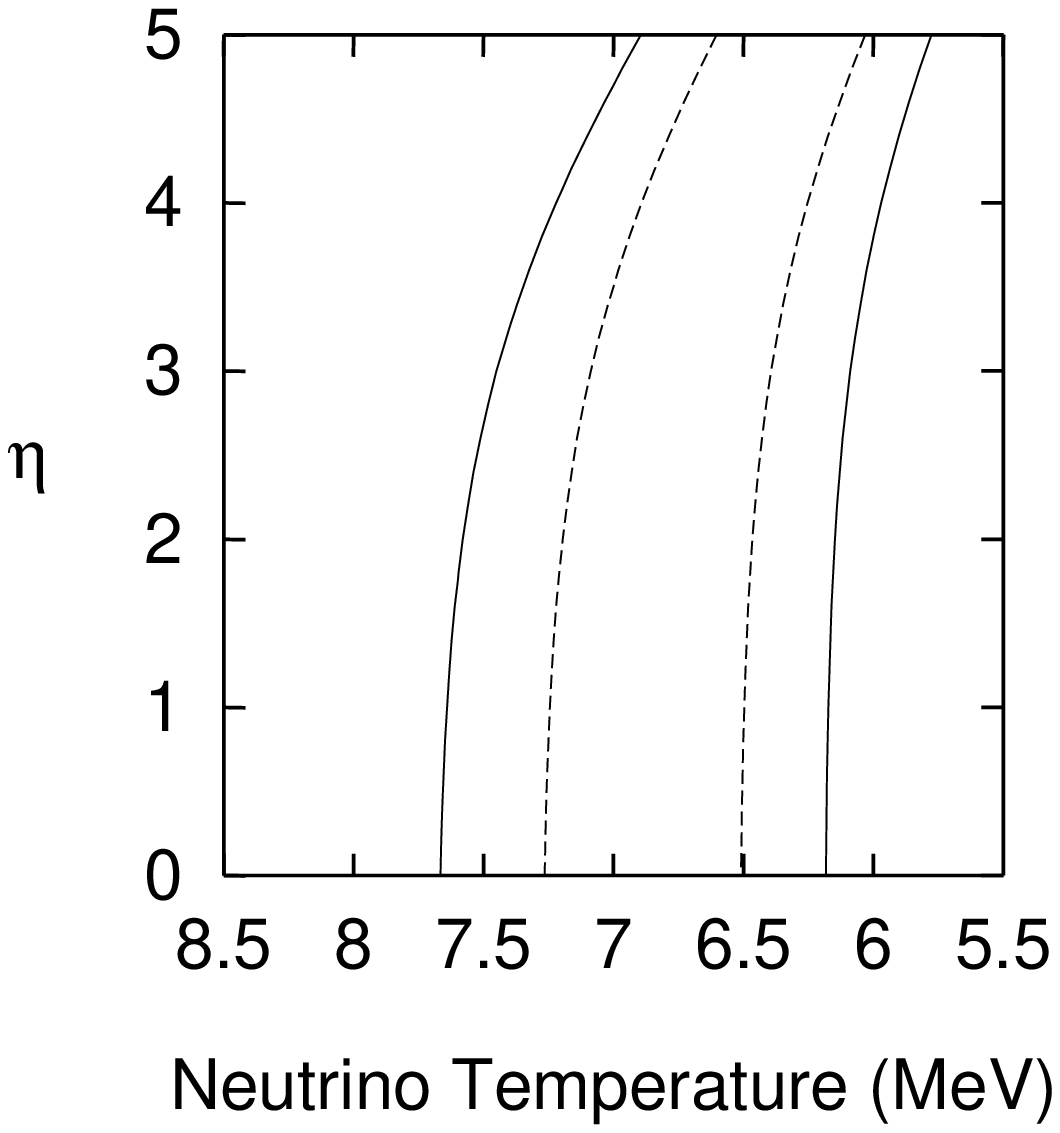}}
\end{minipage}
\end{center}
\caption{\small Contour plot showing the average energies of electrons
emitted from lead with one and two neutrons.  In the plot on the right hand
side, the bands show the $\pm 5\%$ (inner) and $\pm 10\%$ (outer) bands for
an average energy of 20 MeV of an electron emitted with two neutrons. In
the plot on the left hand side, the bands show the $\pm 5\%$ (inner) and
$\pm 10\%$ (outer) bands for an average energy of 23 MeV of an electron
emitted with two neutrons. \label{fig:energies}}
\end{figure}

To show that the calculations of average energies are not very sensitive,
fig.\ \ref{fig:e1} presents the results with the two different forces, SIII
and SKO+. Fig.\ \ref{fig:e2} shows the same plot but with more contours and
only for the force SIII. The contour lines in fig.\ \ref{fig:energies} show
the effects of uncertainties in the calculated average energies.  The
figure shows the $\pm 5\%$ and $\pm 10\%$ regions for an electron spectrum
with an average energy of 20 MeV emitted with one neutron, and for a
spectrum with average energy 23 MeV emitted with two neutrons. Even with an
uncertainty of $\pm 10\%$ (see the last section), the temperature can be
determined to within about 1 MeV.  If the average energies can be
accurately measured, therefore, the temperature of the neutrino
distribution can be determined well. On the other hand, if one considers
only the average energies, a large range of chemical potentials is allowed
(at least for $\eta < 3$), almost independent of how well the average
energy can be measured.

One might ask how much we gain by considering the one- and two-neutron
contours simultaneously.  We could plot the two sets on the same figure,
creating a grid that would in theory allow us to determine both $T$ and
$\eta$.  Unfortunately the contours for the two kinds of events are nearly
parallel except at large $\eta$, so that the such a plot may not provide a
stronger constraint on the flux parameters.

\begin{figure}[h!]
\begin{center}
\begin{minipage}{7.5cm}
\centerline{\includegraphics[angle=0,width=6cm]{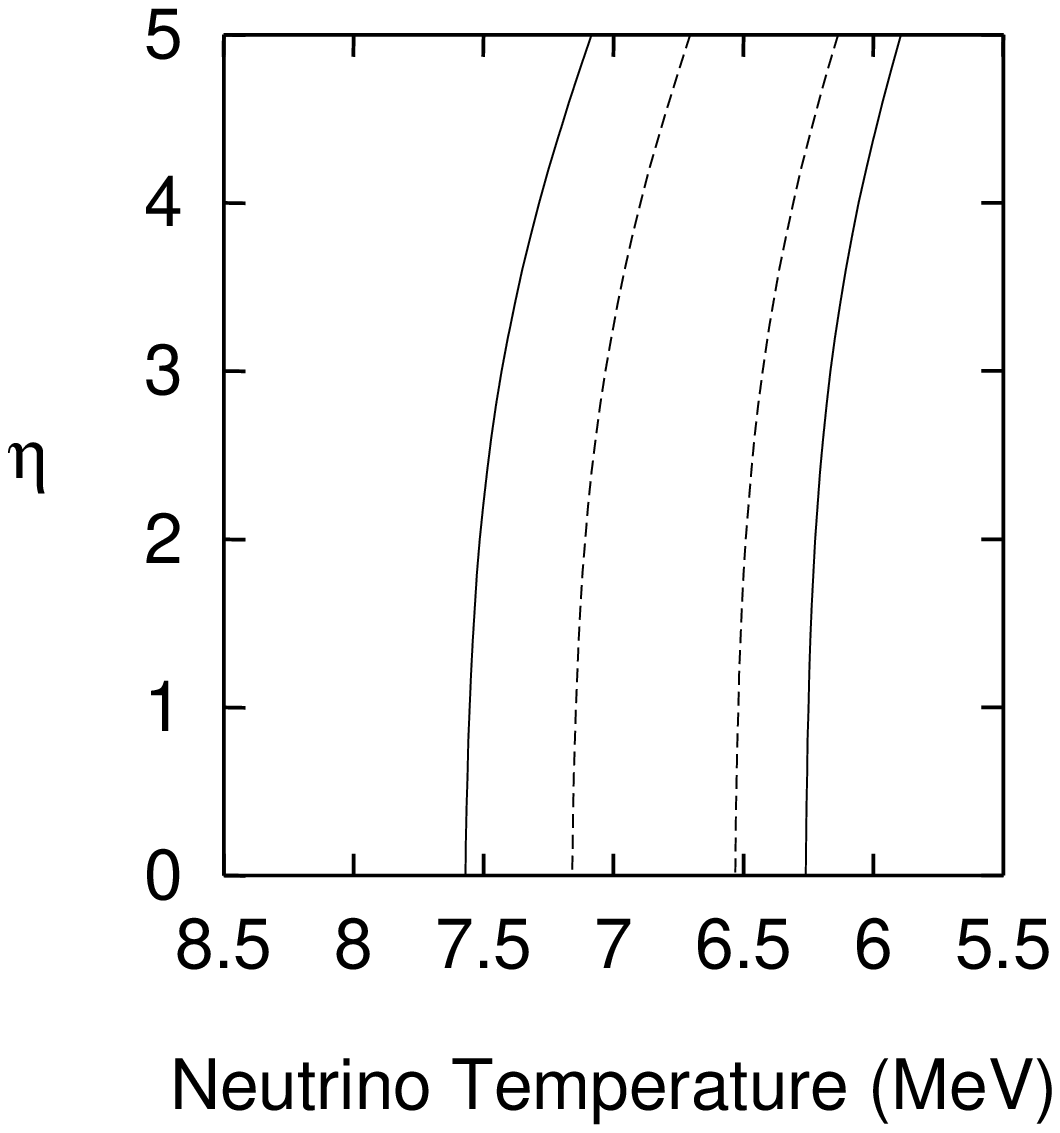}}
\end{minipage}\begin{minipage}{7.5cm}
\centerline{\includegraphics[angle=0,width=6cm]{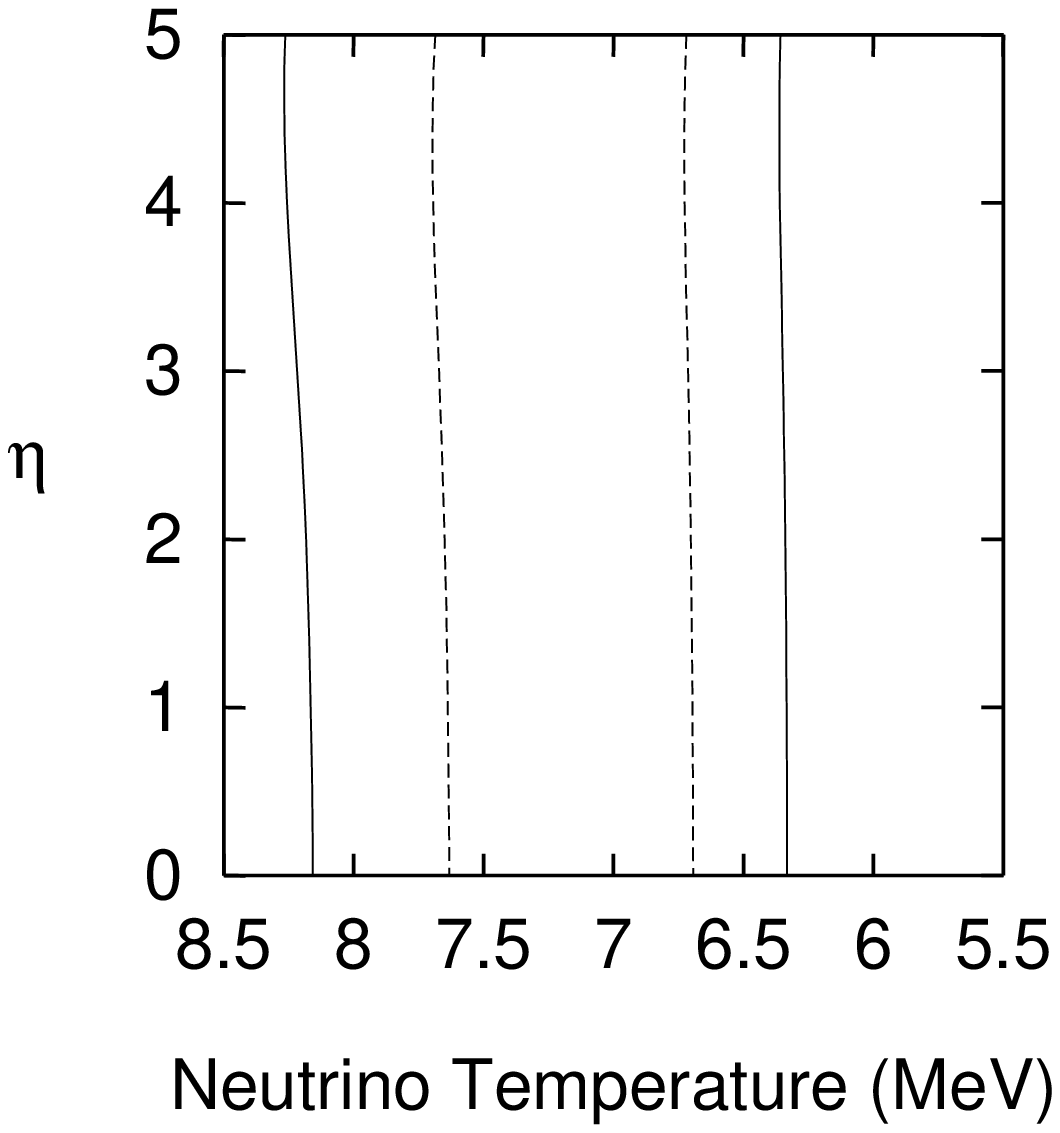}}
\end{minipage}
\end{center}
\caption{\small Same as Fig.\ \ref{fig:energies}, except that here we show
contours  of the width of electron energy distribution for one (left
figure) and two (right figure) neutrons.  We show contours of $\pm 5\%$ and
$\pm 10\%$ for a width of 12 MeV. \label{fig:secondmom}}
\end{figure}

In  fig. \ref{fig:secondmom}, we plot the width of the distribution,
defined as $\langle E_e^2 - \langle E_e \rangle^2 \rangle^{1/2}$ instead of
the average energy.   We use $\pm 5\%$ and $\pm 10\%$ contours for a width
of 12 MeV for both electrons emitted with one neutron and electrons emitted
with two neutrons.  This quantity, too, is not sensitive to $\eta$ unless
$\eta$ is large.  The $\pm 5\%$ and $\pm 10\%$ contours depend, of course,
on what average energy is measured.   Furthermore, a similar analysis, both
here and for the average energies, may be performed with deviations from a
Fermi-Dirac spectrum.

The extent to which the electron spectra are useful for determining the
high-temperature component of the spectrum depends very much on the
contamination from other neutrino sources.  As discussed above, in the case
of partial transformation, the low-temperature component of the spectrum
will contribute very little to the average energies of electrons associated
with two neutrons, but can make almost an MeV difference in the average
energy of electrons associated with one neutron.  Furthermore, if the lead
perchlorate were mixed with water, interactions with Chlorine, Oxygen and
Hydrogen would also occur. Oxygen and Chlorine have high thresholds for the
emission of neutrons, and they also emit protons, so their event rates will
be low. But if the water were 20\% of the detector there would be
approximately 70 events from
\begin{equation}
\bar{\nu}_e + p \rightarrow n + e^+
\end{equation}
(assuming a normal hierarchy with $\sin^2 \theta_{12} \approx 0.75$) per kt
of lead perchlorate. These events will be difficult to
distinguish from the one-neutron charged-current events on lead.

In the case of a normal hierarchy and no earth matter effect, the
antineutrino signal would consist of $\sim \sin^2 \theta_{12} $ of the
original $\bar{\nu}_e$ spectrum and $\cos^2 \theta_{12}$ of the original
$\bar{\nu}_{\mu,\tau}$ spectrum. If a detector such as SuperKamiokande or
KamLAND were operating at the time, it would be able to determine the
$\bar{\nu}_e$ spectrum, making it possible to subtract out that component.
If not, however, one might have to rely on the two-neutron signal.  That by
itself, though, is enough to get a good handle on the temperature of the
high-temperature neutrinos, as discussed above.

We turn finally to the problem of distinguishing  between two likely
neutrino spectra at the detector: a completely hot spectrum associated with
the normal hierarchy and complete conversion at the first resonance, and a
$\sim 75\%$ hot, $\sim 25\%$ 
cold mixture associated with either the inverted hierarchy
or very small $\theta_{13}$. Once the electron energy spectrum has been
used to learn about the hot component of the electron-neutrino energy
spectrum, we can then use the ratio of charged-current to neutral-current
events to get around the problem of not knowing the distance to the
supernova.

\begin{figure}[h!]
\centerline{\includegraphics[angle=-90,width=12cm]{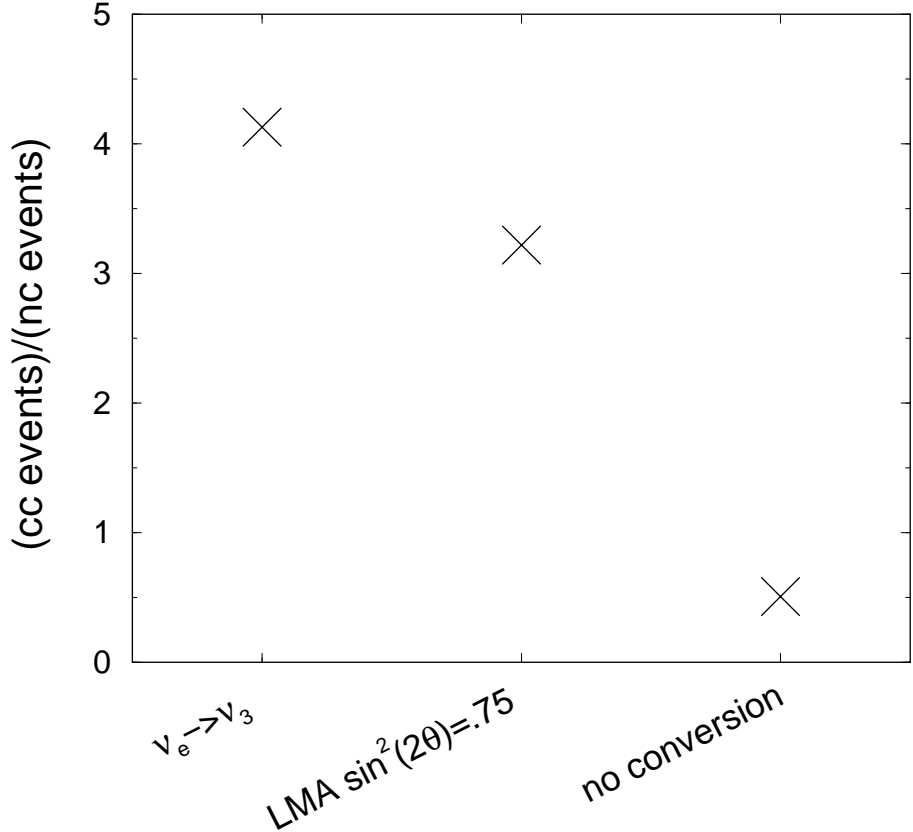}}
\caption{\small Ratio of charged current events to neutral
current events in the case of the various oscillation scenarios.  From left
to right: total conversion at the $\delta m^2_{13}$ resonance, partial
(adiabatic) conversion at the $\delta m^2_{12}$ resonance, and no
conversion. This figure is made assuming T=8.0 MeV $\eta=3$ for the
high-temperature flux, T=5.0 MeV, $\eta=0$ for the original $\bar{\nu}_e$
flux, and T=3 MeV $\eta = 3$ for the original $\nu_e$ flux.
\label{fig:cctonc}}
\end{figure}

Fig. \ref{fig:cctonc} shows the ratio of the number of
charged-current events to the number of neutral-current events (with no
electrons).  
Since in the charged-current reactions the
neutron emission threshold is high, only high energy neutrinos cause events.
Furthermore, the neutral current signal is unaffected by oscillations.
Therefore, the partial-oscillation scenario results in a ratio that is
about 3/4 ( or $\cos^2 \theta_{12}$) 
of what complete oscillations yield.  Clearly, to exploit this
result it is essential to know the reaction cross sections precisely, i.e.\
to within about 10\%.  No calculations, including ours, are that certain
but improvements in the model, e.g.\ through an exact treatment of Coulomb
effects, and (especially) measurements of neutrino/lead cross sections
could get us to that level.

If experimental work on solar neutrinos continues, then at the time a
supernova explodes there may well be a detector in place that can see
electron antineutrinos through the reaction $\bar{\nu}_e + p \rightarrow n
+e^+$. The positron spectrum, which will closely track the incoming
antineutrino distribution, could in principle be used in conjunction with
lead to determine the difference between the normal and inverted
hierarchies.

\section{Conclusions}
\label{sec:conclusions}

In this paper, we have analyzed the properties of a lead-based
supernova-neutrino detector. The expected energy distribution of electrons
emitted with one or two neutrons peaks somewhere in the low 20's of MeV,
depending on the details of the incoming neutrino spectral shape. We expect
a few hundred charged- and neutral-current events per kt of lead.

We have used the Random-Phase-Approximation with effective Skyrme forces to
calculate neutrino-lead cross sections.  Our results  are in agreement with
those of refs.\ \cite{kl} and \cite{volpe}, but lower than the ones in
ref.\ \cite{fhm}, for reasons we understand. We used two different Skyrme
forces, SIII and a parameterization we call SkO+ based on the force SkO';
the average electron energies obtained with these two interactions are
within a few percent of each other. Despite this good agreement, it is
difficult to assess quantitatively the overall systematic uncertainties in
our cross sections. We don't know how much forbidden strength, which is
particularly important in the two-neutron channel, is quenched, or even
much about how it's distributed.  It's possible, however, to improve the
calculations, and additional data from neutrino scattering and electron
scattering would help tremendously.

The spectra of neutrinos reaching the earth can be modified by
oscillations.  We have argued that in most scenarios, the electron energy
spectrum can be used to determine the temperature (and perhaps the
effective chemical potential) of the hot neutrino spectrum, due to
neutrinos that were originally emitted as $\nu_\mu$ or $\nu_\tau$. (We find
that lead has little sensitivity to the cold neutrinos originally emitted
as $\nu_e$, or, in the charged-current channel, to antineutrinos.) With a
$\pm 5\%$ theoretical uncertainty in the average energies of the electrons,
there would be an approximately 0.5 MeV uncertainty in the temperature of
the hot neutrino spectrum.  This doubles if the uncertainty is $\pm 10\%$.
These numbers are small enough so that a lead detector would provide crucial
information about the proto-neutron star.

We discussed the possibility of distinguishing between the complete and
partial neutrino transformation by using the ratio of charged-current to
neutral-current event numbers.  Although the idea works in principle, it
would be hard to use now because of the uncertainty in calculated cross
sections. However, it should be possible to distinguish between the case of no
transformation and either partial or complete transformation.  Since some
degree of transformation is expected, either due to the LMA or
$\theta_{13}$ this ability will provide an important check of our
understanding of neutrino physics.

A measurement of neutrino cross sections on lead is
very important. As discussed above, we need a reasonable degree of
certainty in the average energies to extract information about the incoming
neutrino spectrum. That certainty is even more important if we want to see
whether some or all of the original electron neutrinos have oscillated. The
ability to distinguish between the two would let us use supernovae to learn
something about the size of ${\theta_{13}}$ or discover whether nature has
chosen the normal or inverted hierarchy.

\vspace{1.cm} We wish to thank S. Elliott, J. Beacom, A. Murphy and D. Boyd
 for useful discussions.  Two of us (G.C.M. and C.V.) acknowledge the 
European Centre for Theoretical Studies in Nuclear Physics and Related Areas 
(ECT*). G.C.M. is supported by the U.S. Department of Energy
under grant DE-FG02-02ER41216, and J.E. by the U.S. Department of Energy 
under grant DE-FG02-97ER41019.


\end{document}